\documentclass[prd,article,nofootinbib,twocolumn,preprintnumbers,superscriptaddress,amsmath,amssymb,aps]{revtex4-1}

\usepackage{graphicx}
\usepackage{dcolumn}
\usepackage{bm}
\usepackage[table]{xcolor}
\usepackage{multirow}
\usepackage{comment}
\usepackage{url}

\definecolor{kjkblue}{rgb}{0.39, 0.589, 0.6914}

\DeclareMathAlphabet{\mathpzc}{OT1}{pzc}{m}{it}

\graphicspath{{Figures/}}

\begin{document}

\preprint{FERMILAB-PUB-19-136-T, NUHEP-TH/19-03}

\title{Sub-GeV Atmospheric Neutrinos and CP-Violation in DUNE}

\author{Kevin J. Kelly}
\email{kkelly12@fnal.gov}
\affiliation{Theory Department, Fermi National Accelerator Laboratory, P.O. Box 500, Batavia, IL 60510, USA}
\author{Pedro A. N.  Machado}
\email{pmachado@fnal.gov}
\affiliation{Theory Department, Fermi National Accelerator Laboratory, P.O. Box 500, Batavia, IL 60510, USA}
\author{Ivan Martinez-Soler}
\email{ivan.martinezsoler@northwestern.edu}
\affiliation{Theory Department, Fermi National Accelerator Laboratory, P.O. Box 500, Batavia, IL 60510, USA}
\affiliation{Department of Physics \& Astronomy, Northwestern University, Evanston, IL 60208, USA}
\affiliation{Colegio de F\'isica Fundamental e Interdisciplinaria de las Am\'ericas (COFI), 254 Norzagaray street, San Juan, Puerto Rico 00901.} 
\author{Stephen J. Parke}
\email{parke@fnal.gov}
\affiliation{Theory Department, Fermi National Accelerator Laboratory, P.O. Box 500, Batavia, IL 60510, USA}
\author{Yuber F. Perez-Gonzalez}
\email{yfperezg@northwestern.edu}
\affiliation{Theory Department, Fermi National Accelerator Laboratory, P.O. Box 500, Batavia, IL 60510, USA}
\affiliation{Department of Physics \& Astronomy, Northwestern University, Evanston, IL 60208, USA}
\affiliation{Colegio de F\'isica Fundamental e Interdisciplinaria de las Am\'ericas (COFI), 254 Norzagaray street, San Juan, Puerto Rico 00901.}

\date{\today}

\begin{abstract}
    We propose to use the unique event topology and reconstruction capabilities of liquid argon time projection chambers to study sub-GeV atmospheric neutrinos.
    The detection of low energy recoiled protons in DUNE allows for a determination of the leptonic $CP$-violating phase independent from the accelerator neutrino measurement.
    Our findings indicate that this analysis can exclude several values of $\delta_{CP}$ beyond the $3\sigma$ level.
    Moreover, the determination of the sub-GeV atmospheric neutrino flux will have important consequences in the detection of diffuse supernova neutrinos and in dark matter experiments.
\end{abstract}

\pacs{Valid PACS appear here}
\maketitle

\section{Introduction}

Atmospheric neutrinos, produced by cosmic-ray interactions in the Earth's atmosphere, have played a crucial role in the discovery of neutrino oscillations~\cite{Fukuda:1998mi}, the only evidence of non-zero neutrino masses~\cite{Kajita:2016cak, McDonald:2016ixn}. Even now, atmospheric neutrinos contribute significantly to our understanding of neutrino oscillations and mixing in the lepton sector~\cite{Abe:2017aap, Aartsen:2017nmd}.
In this Letter, we are particularly interested in such neutrinos with energies in the 100~MeV to 1~GeV region, i.e. sub-GeV atmospheric neutrinos.
The oscillation phenomenology of this sample is exceptionally rich~\cite{Barger:1998ed, Friedland:2004ah, Huber:2005ep, Akhmedov:2008qt, Mena:2008rh, Peres:2009xe, FernandezMartinez:2010am, Barger:2012fx, Akhmedov:2012ah, Winter:2013ema}. 
The physical reason behind this is twofold. 
First, for baselines comparable to the Earth's radius, oscillation of sub-GeV neutrinos are strongly affected by both solar and atmospheric mass splittings. 
Second, the broad energy spectrum and large matter effects induced by the Earth's matter density profile lead to non-trivial oscillation effects, namely MSW~\cite{Wolfenstein:1977ue, Mikheev:1986gs} and parametric~\cite{Akhmedov:1988kd, Krastev:1989ix} resonances.
Compared to long baseline accelerator neutrinos, the effects on oscillation probabilities of the leptonic  $CP$-violating phase $\delta_{CP}$ is much more pronounced in sub-GeV atmospheric neutrinos, and therefore, a  measurement of their oscillation pattern can yield important new information on $\delta_{CP}$.

At present, only Cherenkov light detectors like Super-Kamiokande and IceCube are large enough to have significant sensitivity to the broad spectrum of atmospheric neutrinos. 
Nevertheless, a precise measurement of sub-GeV neutrinos is still lacking.
The difficulty in studying these neutrinos is related to the event reconstruction which is very challenging at these low energies.
When a sub-GeV neutrino scatters on a nucleon via a charged current interaction, it produces a charged lepton and recoils the nucleon isospin partner, for instance $\nu_e n\to e^- p^+$. 
However, protons with less than about 1.4~GeV of kinetic energy do not emit any Cherenkov light in water, and thus are as invisible as neutrons in these detectors.

In neutrino scattering events, the kinematics of the outgoing lepton bears correlation with the neutrino energy and direction. 
Nevertheless, at these low energies much of such correlation is lost, and there is a large spread in outgoing lepton momentum and angle. 
On top of that, the lack of charge identification impedes the separation of events originated from neutrinos or antineutrinos, concealing $CP$-violating effects.
All this results in very poor reconstruction of the neutrino energy, direction and flavor (between $\nu$ and $\bar\nu$), making  the use of sub-GeV atmospheric neutrinos to probe $CP$-violation in Cherenkov detectors impractical, unless  detectors are gigantic, at the multi-megaton scale~\cite{Razzaque:2014vba}. 

In liquid argon time projection chambers (LArTPCs), the situation is completely different.
The LArTPC technology allows for excellent reconstruction of neutrino event topologies by detecting the tracks of all charged particles and identifying them by topology and energy loss.
Recently it was shown that protons with kinetic energy above 21~MeV can be efficiently identified in the ArgoNeut experiment~\cite{Palamara:2016uqu}. Their three-momenta can still be reconstructed with good resolution, which will allow for a pioneering measurement of sub-GeV neutrino energies and angles.
Besides, the capability of detecting these protons allows for statistical separation between sub-GeV neutrinos and antineutrinos, since the former is significantly more likely to kick out a single proton from Argon than the latter~\cite{Palamara:2016uqu}. 
Together with the fact that the cross section for neutrinos is about a factor of 2 larger than the one for antineutrinos, one expects that events with one  lepton, one proton, and no pions (CC-$1p0\pi$) are ``neutrino-rich,'' while events with only an outgoing lepton (CC-$0p0\pi$) are somewhat ``antineutrino-rich.''

In this Letter, we propose to use the unique event reconstruction capabilities of LArTPC to estimate how the future Deep Underground Neutrino Experiment (DUNE)~\cite{Acciarri:2015uup} will be able to measure  sub-GeV atmospheric neutrinos and extract  information on $\delta_{CP}$ complementary to the accelerator neutrino program.
In addition to $\delta_{CP}$, the study of sub-GeV atmospheric neutrinos will have major impact in the determination of the diffuse supernova neutrino background and the neutrino background in dark matter direct detection experiments, as well as in searches for new physics in the neutrino sector.

\section{Physics with Low-Energy Atmospheric Neutrinos}\label{sec:physics}

In general terms, neutrino oscillations are driven by a phase
$\propto (\Delta m_{ij}^2/{\rm eV}^2) (L/{\rm km}) ({\rm GeV}/E)$, where  $L$ is the distance traveled between neutrino production and detection, $E$ is the neutrino energy, and $\Delta m_{ij}^2 \equiv m_i^2 - m_j^2$ is the squared mass splitting. 
When $E \gtrsim 1$ GeV, oscillations are induced largely by the aptly-named atmospheric mass splitting $|\Delta m_{31}^2| \simeq 2.5 \times 10^{-3}$~eV$^2$~\cite{NOvA:2018gge, Abe:2018wpn}, and they develop over scales $L \sim \mathcal{O}(R_E)$, the radius of the Earth.
Oscillations of atmospheric neutrinos with energies $100~{\rm MeV}< E < 1~{\rm GeV}$, are governed by both the atmospheric mass splitting and the smaller solar mass splitting, $\Delta m_{21}^2 \simeq 7.4\times 10^{-5}$ eV$^2$~\cite{Gando:2013nba, Cleveland:1998nv, Abdurashitov:2009tn, Bellini:2011rx, Hosaka:2005um, Aharmim:2011vm}.
In what follows, we will consider two aspects of major significance to our analysis, $CP$ violation and matter effects.
We adopt the usual parametrization for neutrino mixing~\cite{Tanabashi:2018oca}.
To set convention, we define the zenith angle such that $\cos\theta_z=-1$ corresponds to neutrinos coming from directly below the detector, while $\cos\theta_z=0$ indicates the horizon direction.

First we discuss the effects of $\delta_{CP}$ in oscillations of sub-GeV neutrinos. 
In vacuum, for simplicity, the $CP$-violating term in  neutrino oscillation probability is given by~\cite{Denton:2016wmg}
\begin{equation}
  P_{CP}=-8J_r\sin\delta_{CP}\sin\Delta_{21}\sin\Delta_{31}\sin\Delta_{32},
\end{equation}
which includes the Jarlskog invariant~\cite{Jarlskog:1985ht, Jarlskog:1985cw} $J_r\sin\delta_{CP}$ (in our convention) and  $\Delta_{ij}\equiv\Delta m^2_{ij}L/4E$ are the oscillation phases.
Oscillations of beam neutrinos probe the atmospheric splitting $\Delta_{31}\sim\mathcal{O}(1)$, while $\Delta_{21} \ll 1$. 
There, the $CP$ term is suppressed by $\Delta m^2_{21}/\Delta m^2_{31}\times\pi/2\sim 1/20$ due to the fact that  oscillations driven by $\Delta m^2_{21}$ do not have time to develop. This yields
$P_{CP}\simeq -0.4J_r \sin\delta_{CP}\sin\Delta_{31}\sin\Delta_{32}$.
Sub-GeV atmospheric neutrino oscillations, on the other hand, probe the solar splitting. 
In this case, the oscillations driven by $\Delta m^2_{31,32}$ are fast and average out. 
The resulting factor is just $1/2$, leading to a much larger $CP$-violating term relative to beam neutrinos, namely $P_{CP}\simeq-4J_r^m\sin\delta_{CP}\sin\Delta_{21}$ with $\Delta_{21}\sim\mathcal{O}(1)$.

\begin{figure}[t]
  \centering
  \includegraphics[width=\linewidth]{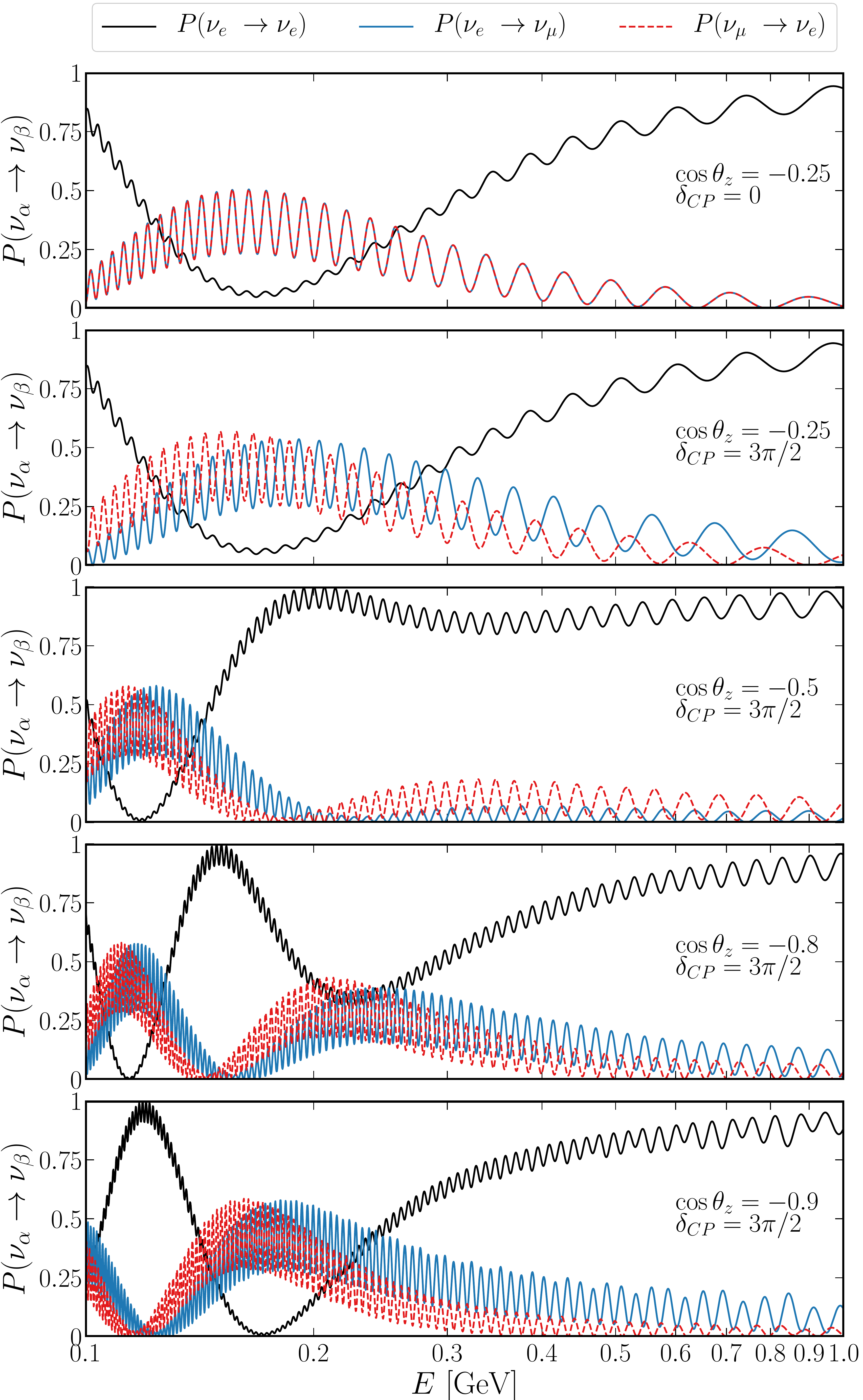}
  \caption{Oscillation probabilities for $\nu_e\to\nu_e$ (black), $\nu_e\to\nu_\mu$ (blue), and $\nu_\mu\to\nu_e$ (dashed red) for various values of the zenith angle $\cos\theta_z$ as indicated, and $\delta_{CP}=0$ (top panel) or $\delta_{CP}=3\pi/2$ (all other panels). Earth's matter profile was implemented using the PREM model~\cite{Dziewonski:1981xy}. In the upper panel, the red and blue lines lie on top of each other. \label{fig:prob-dcp}}
\end{figure}

In Fig.~\ref{fig:prob-dcp}, we  present several oscillation probability curves\footnote{See \url{https://imgur.com/HoWUniu} for an animation of how $\delta_{CP}$ changes oscillation probabilities as a function of zenith angle and neutrino energy.} as function of neutrino energy for $\nu_e\to\nu_e$ (black) $\nu_e\to\nu_\mu$ (blue) and $\nu_\mu\to\nu_e$ (dashed red), various zenith angles and $\delta_{CP}=0$ (upper panel) or $\delta_{CP}=3\pi/2$ (all other panels).
Focusing on the blue and red curves in the first two  panels, we observe a large effect of $\delta_{CP}$, as a non-zero value of this phase splits out the $\nu_\mu\to\nu_e$ and $\nu_e\to\nu_\mu$ appearance probabilities. 
This large effect will enhance the sensitivity of sub-GeV atmospheric neutrinos to $\delta_{CP}$.
The second feature that stands out is the impact of different zenith angles, which is related to matter effects.
We turn our attention to them now.

The second crucial feature of sub-GeV atmospheric neutrino oscillations are matter effects. Interactions with matter in the Earth, specifically in the dense mantle and core, may significantly modify neutrino oscillations.
These effects are quite rich and have been studied in great depth~\cite{Barger:1998ed, Friedland:2004ah, Huber:2005ep, Akhmedov:2008qt, Mena:2008rh, Peres:2009xe, FernandezMartinez:2010am, Barger:2012fx, Akhmedov:2012ah, Winter:2013ema}. 
Here we restrict ourselves to review some oscillation aspects and provide a few examples.

Up-going atmospheric neutrinos that traverse the Earth may go through an MSW resonance~\cite{Wolfenstein:1977ue, Mikheev:1986gs} in the solar sector, maximizing oscillations between $\nu_e$ and $\nu_{\mu,\tau}$, when
\begin{equation}
    \Delta m^2_{21} \cos\theta_{12} = 2 \sqrt{2}E G_F n_e,
\end{equation}
where $G_F$ is the Fermi constant, and $n_e$ is the electron number density. 
In the solar sector, the MSW resonance happens only for neutrinos, not for antineutrinos, as observed in oscillation of neutrinos produced in the Sun.
We will focus on the $\nu_e\to\nu_e$ oscillation dependence on the zenith angle, shown as black curves in the different panels of Fig.~\ref{fig:prob-dcp}.
In the crust (first or second panels, $-0.44<\cos\theta_z$), mantle  (third and fourth panels, $-0.84<\cos\theta_z<-0.44$) and core (bottom panel, $\cos\theta_z<-0.84$), the MSW resonant energies are found to be around 180, 130, and 50 MeV, respectively.
Although this energy in Earth's core is below 100~MeV, another type of resonance occurs about $E\sim 170$~MeV, a parametric resonance~\cite{Akhmedov:1988kd, Abe:2018wpn, Krastev:1989ix}. 
A parametric resonance happens when changes to the matter density profile occur on the same scale as the neutrino oscillation length.
The phenomenon is analogous to a resonant spring oscillator.
The fast oscillations (which begin to slow down as $E\to 1$ GeV) are induced by the atmospheric mass splitting.
Note that, due to the near-maximal value of $\theta_{23}$, $\nu_e$ oscillates approximately equally into $\nu_\mu$ and $\nu_\tau$.

The $CP$-violating and matter effects displayed in Fig.~\ref{fig:prob-dcp} show that the $\delta_{CP}$ effect is broad in neutrino energy, but there are large variations of oscillation curves for different zenith angles.
Therefore, the precise reconstruction of the neutrino energy will not be as important as the determination of the incoming neutrino direction for the measurement of $\delta_{CP}$.
LArTPCs have excellent energy resolution and tracking reconstruction, and hence the incoming neutrino direction may be determined by considering the full event topology in charged current quasi-elastic events, $\nu_\ell n\to\ell^-p^+$.
In the next Section, we will discuss the details of our simulation of sub-GeV atmospheric neutrinos and how we take the nuclear physics effects into account.

\section{Simulation Details}
\label{sec:Simulation}

To simulate the atmospheric neutrino flux at sub-GeV energies, we use Ref.~\cite{Honda:2015fha}. 
The atmospheric neutrinos  flux for a given flavor is parametrized by
\begin{equation}
    \Phi_\alpha(E)=\Phi_{\alpha,0}\,f_\alpha(E)\left(\frac{E}{E_0}\right)^{\delta},
\end{equation}
where $f_\alpha(E)$ gives the shape of  the neutrino energy spectrum for each flavor (obtained from simulations, see Ref.~\cite{Honda:2015fha}), $\Phi_{\alpha,0}$ is the  normalization of the flavor $\alpha$  ($\nu_e,\nu_\mu,\bar\nu_e,\bar\nu_\mu$), $E_0$ is an arbitrary reference energy, and $\delta$ accounts for spectral distortions.
To account for unknowns on the meson production in the atmosphere, we consider systematic uncertainties on the following quantities: overall normalization ($40\%)$; the ratio $r_e$ between $\nu_e$ and $\nu_\mu$ fluxes  (5\%); the ratio $r_\nu$ between  neutrinos  and antineutrinos  fluxes (2\%); and the spectral distortion parameter $\delta$ with $0.2$ absolute uncertainty.

\begin{figure}[t]
  \centering 
  \includegraphics[width=0.45\linewidth]{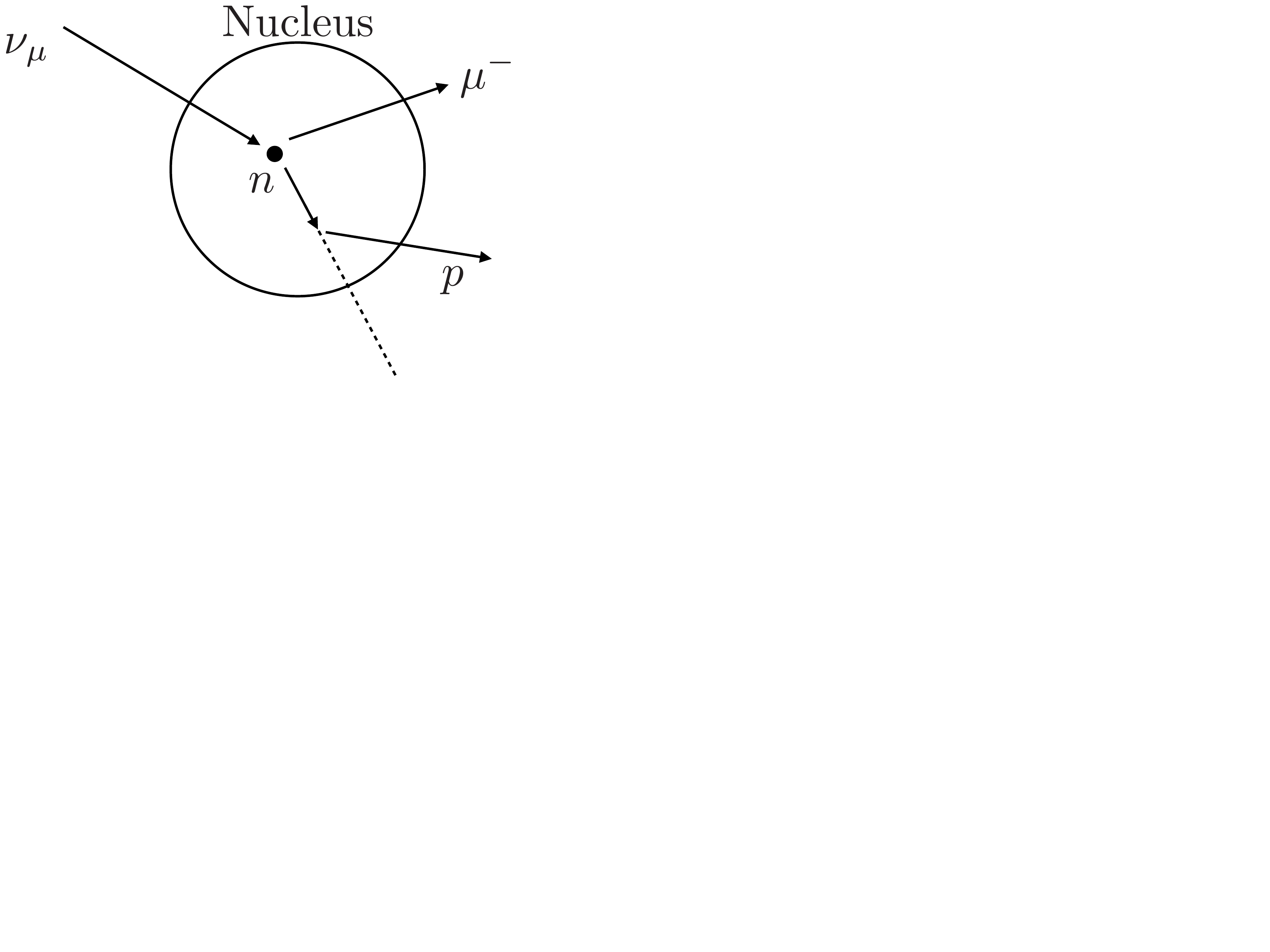}
  \caption{Pictorial representation of an intra-nuclear cascade.  \label{fig:cascade}}
\end{figure}

Neutrino events in the DUNE detector will be classified by topology. 
We consider events with a charged lepton (electrons or muons) and up to 2 outgoing protons and no pions, namely CC-$Np0\pi$ ($N = 0, 1, 2$).
The interaction of neutrinos scattering  on argon  was modeled with the  NuWro event generator~\cite{Golan:2012rfa}.
This is an important step as recoiled nucleons may re-interact still inside the nucleus, a process typically referred to as final state interactions or intra-nuclear cascades. 
A pictorial representation of intra-nuclear cascades is shown in Fig.~\ref{fig:cascade}.
To account for detector response, a cut on the minimum proton kinetic energy of 30~MeV was implemented~\cite{Acciarri:2015uup}. Momentum resolutions of $5\%$, $5\%$ and $10\%$ for electrons, muons and protons were assumed~\cite{Acciarri:2014gev} as well as conservative angular resolutions of $5^\circ$, $5^\circ$ and $10^\circ$, respectively~\cite{Alion:2016uaj}. 
We have checked that these angular and energy resolutions are not limiting factors in our results.

Incoming neutrino energy and direction were estimated by reconstructing the energy and direction of all outgoing charged particles, summing up their four momenta, and subtracting the four momenta of initial nucleons assuming they were at rest.
For example, in a CC-$2p0\pi$ event the reconstructed neutrino energy would be given by $E_\nu^\textrm{rec}=E_{\ell}+K_p^{(1)}+K_p^{(2)}$, where $K_p$ indicates the proton kinetic energy.
Besides the imperfect detector response, intra-nuclear cascades as well as outgoing neutrons (which we consider to always go undetected) can affect the neutrino energy and direction reconstruction.
We find that the largest contribution to energy and direction mis-construction arrives from intra-nuclear cascades~\cite{Golan:2012rfa}.
A similar technique was proposed in Ref.~\cite{Rott:2016mzs, Rott:2019stu} to improve the DUNE sensitivity for dark matter annihilation in the Sun using pointing.

To evaluate the experimental sensitivity to $\delta_{CP}$, we have calculated the oscillation probabilities for $-1 \leq \cos{\theta_z} \leq 1$ 
and $100$ MeV $\leq E_\nu \leq 1$ GeV,
assuming the PREM Earth Density Model~\cite{Dziewonski:1981xy} and fixing all oscillation parameters but $\delta_{CP}$ to $(\sin^2\theta_{12},\sin^2\theta_{13},\sin^2\theta_{23})=(0.310,\,0.02241,\,0.580)$, $\Delta m^2_{21}=7.39\times10^{-5}$~eV$^2$, and $\Delta m_{31}^2=+2.525\times 10^{-3}$~eV$^2$, see Ref.~\cite{Esteban:2018azc}. 
We assume an exposure of 400~kton-year. For these values of the oscillation parameters, we expect $\mathcal{O}(4000)$ $\nu_e$ events, $\mathcal{O}(5000)$ $\nu_\mu$ events, and $\mathcal{O}(1000)$ $\overline{\nu}_e$ and $\overline{\nu}_\mu$ events each. The majority of the $\nu$ events are of the CC-$1p0\pi$ topology, where the majority of the $\overline{\nu}$ are CC-$0p0\pi$.

Predicted number of events as a function of reconstructed energy/zenith angle are used to calculate a $\chi^2$ test statistics for each distinct final-state event topology.
No statistical charge identification technique was used, but the detector is assumed to have perfect $\mu$-$e$ separation. 
The sensitivity to $\delta_{CP}$, presented in the following sections, comes from combining of all these event topologies and marginalizing the test statistics over the  systematic uncertainties discussed above.

\section{Discussion}
\label{sec:Results}

The sensitivity to $\delta_{CP}$ for an input value of $\delta_{CP}=3\pi/2$ is shown in Fig.~3 assuming a 400~kton-year exposure. 
The individual $\Delta\chi^2$ contribution for each topology is shown, as well as the combined fit.
A remarkable sensitivity to $\delta_{CP}$ may be achieved, allowing for excluding regions of the parameter space beyond the $3\sigma$ level.

\begin{figure}[t]
  \hspace{-0.4cm}\centering 
  \includegraphics[width=1.02\linewidth]{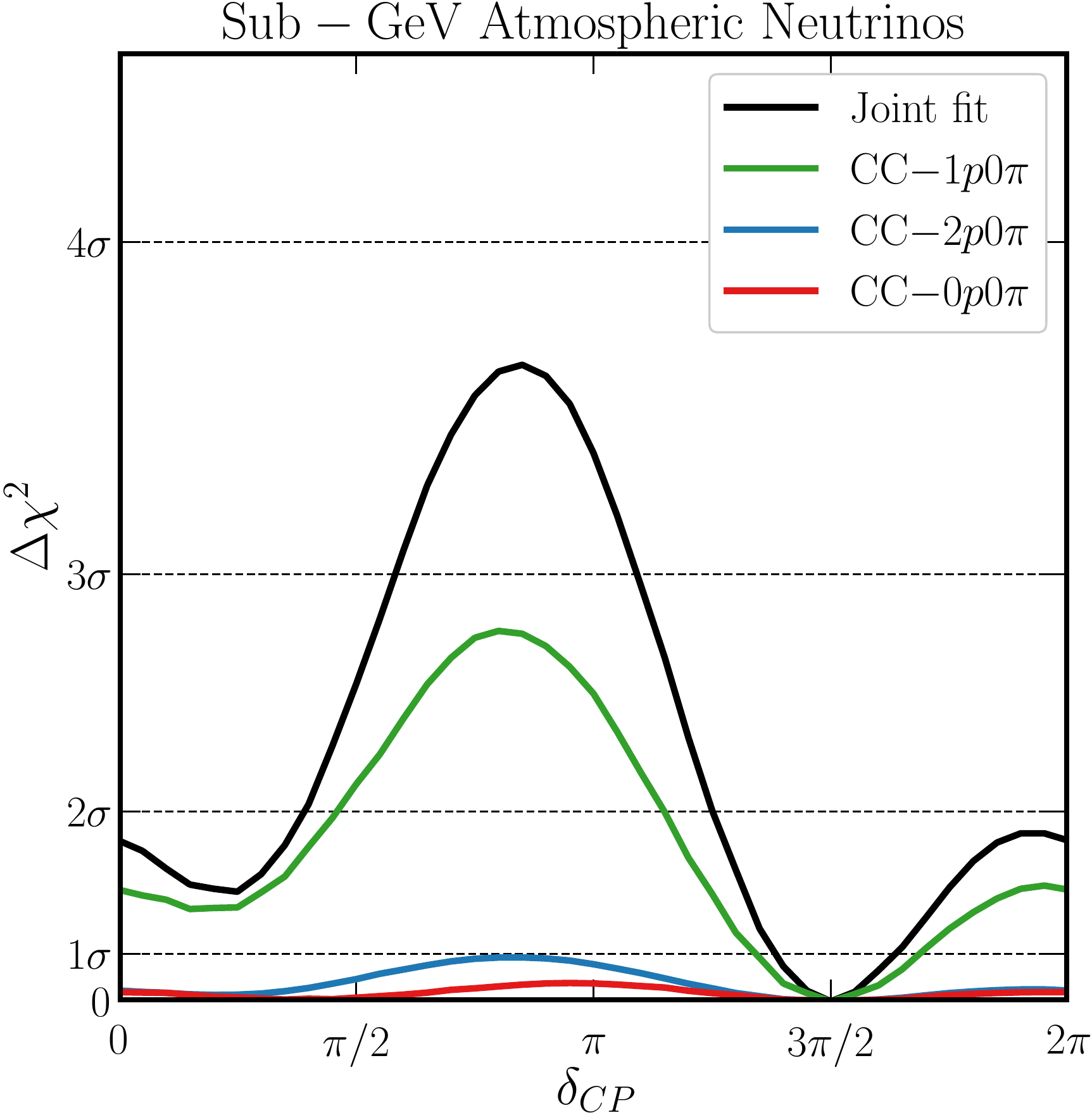}
    \caption{DUNE sensitivity to the leptonic CP violating phase $\delta_{CP}$ using sub-GeV atmospheric neutrinos, for an input value $\delta_{CP}=3\pi/2$.
  An exposure of 400~kton-year was assumed. \label{fig:chisq}}
\end{figure}

Several factors contribute to this sensitivity.
As already discussed, the $CP$ violation effect for sub-GeV atmospheric neutrinos is a sizable effect, an order of magnitude larger than the corresponding one for beam neutrinos.
To observe $CP$ violation, one should be able to independently measure oscillations of neutrinos and antineutrinos and/or the time-conjugated channels $\nu_\mu\to\nu_e$ and $\nu_e\to\nu_\mu$.
The event topology reconstruction in LArTPCs allows for counting the number of protons in the final state.
At these low energies, a neutrino interaction is more likely to kick out a proton from a nucleus than an antineutrino interaction, and vice-versa for neutrons -- therefore, the CC-$1p0\pi$ sample is neutrino rich while CC-$0p0\pi$ is antineutrino rich. 
Combining these two samples allows for measuring, statistically, the flux of $\nu$ and $\bar\nu$ from the atmosphere.
Besides, the incoming neutrino zenith direction has a typical spread, mainly due to intra-nuclear cascades, between $\Delta\theta\sim20^\circ-30^\circ$ using our reconstruction technique, except for the CC-$0p0\pi$ topology which has $\Delta\theta\sim50^\circ$.
This allows to infer the neutrino direction fairly well, disentangling the rich oscillation effects, discussed in Sec.~\ref{sec:physics}, for different baselines.
These aspects indicate a synergy between each distinct topology, as it can be seen in Fig.~\ref{fig:chisq}: the sum of the  individual $\Delta\chi^2$ contributions for each topology is significantly below the combined sensitivity.

\begin{figure}[t]
  \hspace{-0.4cm}\centering 
  \includegraphics[width=1.02\linewidth]{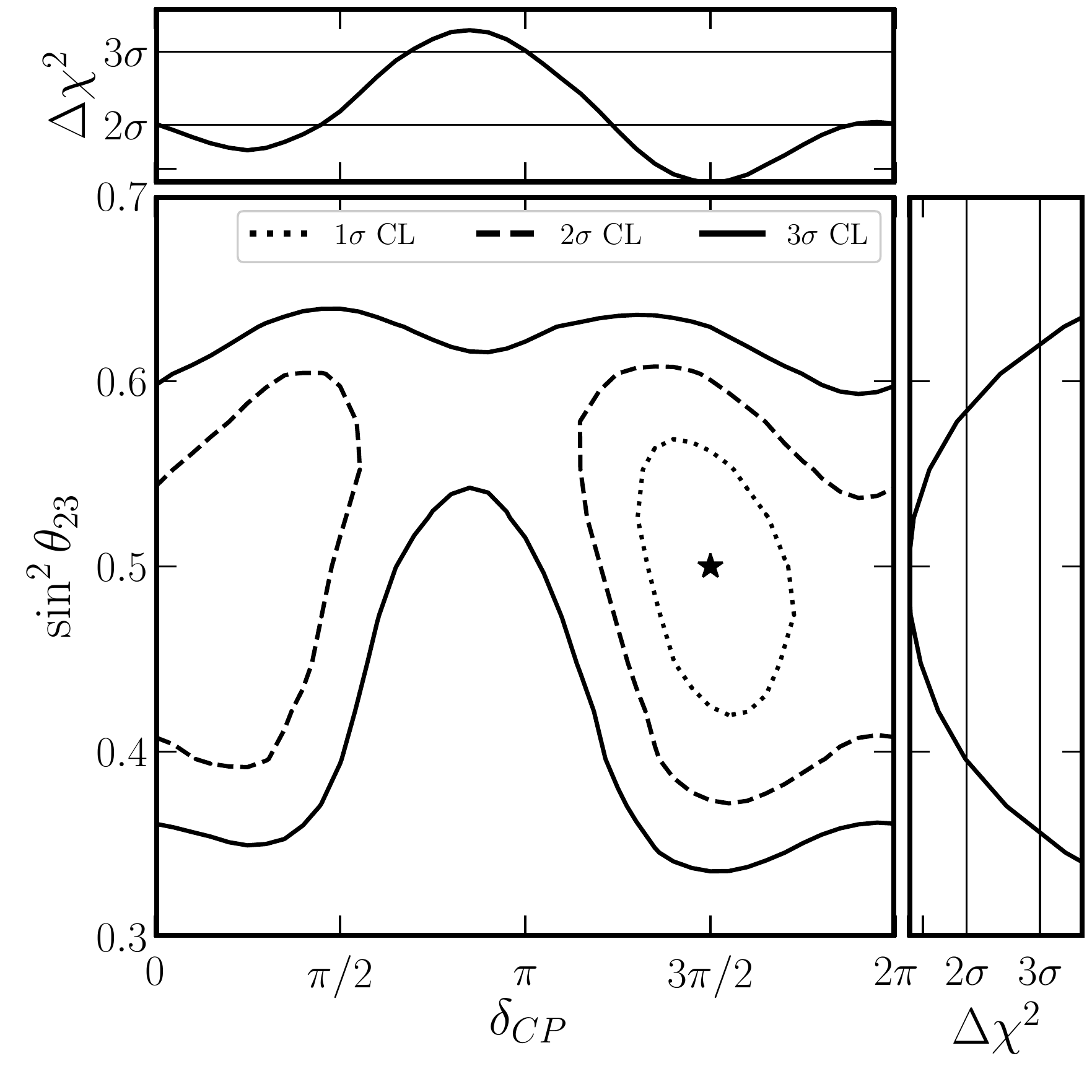}
    \caption{DUNE sensitivity to the leptonic CP violating phase $\delta_{CP}$ and $\sin^2\theta_{23}$ using sub-GeV atmospheric neutrinos. The input values for the two parameters are shown as a star.
    An exposure of 400~kton-year was assumed. \label{fig:chisq2D}}
\end{figure}

With respect to the systematics, we have found that DUNE constrains the pull parameters beyond the uncertainties adopted here, evidencing that the experimental sensitivity is not induced by any prior uncertainty on the atmospheric fluxes, and therefore is quite robust.
We have found that the overall normalization of sub-GeV atmospheric neutrinos can be constrained at the 2\% level, while $r_e$ and $r_{\nu}$ are determined at the 2\% and 1\% level, respectively, and the spectral index of the flux  is constrained as $\delta = \pm 0.02$.
This is an important finding for the detection of the neutrino flux from  diffuse supernovae, as low energy atmospheric neutrinos are among the largest backgrounds to this measurement~\cite{Beacom:2010kk}.
Besides, low energy atmospheric neutrinos are an important component of the neutrino floor, the neutrino-nucleus coherent scattering background in dark matter direct detection~\cite{Gutlein:2010tq}.

The sensitivity to $\delta_{CP}$ obtained here, though not as powerful as   the one obtained with beam neutrinos, is quite competitive.
This will provide a very important cross check for the determination of $\delta_{CP}$ involving energies and baselines very different from those in beam neutrinos.
The experimental sensitivity to $\delta_{CP}$ is limited by statistics and the reconstruction of the original neutrino zenith angle.
We expect that the addition of multi-GeV atmospheric neutrinos will further improve the $CP$ sensitivity, but we leave this study for future work.
Moreover, this distinct configuration will also boost new physics searches such as effective non-standard neutrino interactions and sterile neutrino scenarios.

As a side remark, since the MSW resonance is related to $\Delta m^2_{21}$, and oscillations here are driven by $\theta_{12}$ and $\theta_{23}$, this sample may also be used to constrain other oscillation parameters. 
In Fig.~\ref{fig:chisq2D}, the experimental sensitivity in the $\delta_{CP}\times\sin^2\theta_{23}$ plane (again, after marginalizing over systematic uncertainties) is shown for DUNE sub-GeV atmospheric neutrino analysis, again for 400~kton-year exposure for  $\sin^2\theta_{23}=0.5$ and $\delta_{CP}=3\pi/2$ as input.
Even assuming no prior knowledge on $\sin^2\theta_{23}$ as done here, a  determination of $\sin^2\theta_{23}=0.5\pm0.1$ can be achieved for maximal mixing at $1\sigma$, with little impact on the $\delta_{CP}$ sensitivity.
We leave a detailed analysis to future work, but preliminary results show that the most competitive constraints coming from sub-GeV atmospheric neutrinos are indeed on $\delta_{CP}$. Besides, a similar analysis could be performed for the JUNO experiment~\cite{An:2015jdp}, as JUNO is also capable of identifying low energy charged particles.

Finally, one could wonder what is the role of neutrino interaction uncertainties in this analysis.
We will investigate that in detail in a forthcoming publication, but we would call the attention to the fact that the Short-Baseline Neutrino program~\cite{Machado:2019oxb} at Fermilab and DUNE will explore the sub-GeV region with high statistics.
In fact, the DUNE-PRISM concept~\cite{DUNE-PRISM}, a movable near detector to probe off-axis neutrinos, could greatly enhance our knowledge of neutrino-argon interactions at sub-GeV scales if the near detector hall allows for an off-axis distance of at least $\sim25$~meters.
This would allow for a pioneering data-driven analysis of $CP$ violation using sub-GeV atmospheric neutrinos in DUNE, fully exploring the unique capabilities of liquid argon time projection chambers.

\section{Conclusions}

We have proposed to use the unique capabilities of liquid argon time projection chambers to explore the physics of sub-GeV atmospheric neutrinos.
By detecting low energy charged particles, the direction and energy of incoming neutrinos can be inferred, furnishing LArTPCs with a unique opportunity to probe $CP$ violation with low energy atmospheric neutrinos.
We have shown, with a detailed simulation, that DUNE's sensitivity to the  $CP$ phase from this atmospheric sample is competitive, possibly ruling out regions of the parameter space beyond the $3\sigma$ level, and will provide a remarkable cross check of the $CP$ phase determination with beam neutrinos.
This measurement will have significant consequences in the possible discovery of neutrinos from diffuse supernovae, and in the determination of the neutrino floor in dark matter direct detection experiments, besides opening novel possibilities to probe new physics scenarios.
We also highlight the possibility of performing a data-driven analysis using inputs from highly off-axis DUNE-PRISM measurements on neutrino-argon interactions.

\acknowledgments

We thank Alvaro Hernandez-Cabezudo, Shirley Li, Bryce Littlejohn, Alberto Marchionni, Ornella Palamara and Sam Zeller for useful discussions.
Fermilab is operated by the Fermi Research Alliance, LLC under contract No. DE-AC02-07CH11359 with the United States Department of Energy.  This project has received funding/support from the European Unions Horizon 2020 research and innovation programme under the
Marie Sklodowska-Curie grant agreement No 690575 and
No 674896. IMS and YPG acknowledge travel support from the Colegio de Fisica Fundamental e Interdisciplinaria de las Americas (COFI).


\bibliography{references}

\end{document}